\documentclass[aps,prd,twocolumn,showpacs,amsmath,amssymb]{revtex4-1}
\usepackage{amsmath}
\usepackage{graphicx}
\usepackage{subfigure}
\usepackage{epstopdf}
\usepackage{color}
\usepackage{multirow}
\usepackage{setspace}
\usepackage{overpic}
\usepackage{amssymb}
\usepackage{makecell}
\usepackage[bookmarksnumbered, pdfstartview=FitH,colorlinks,urlcolor=blue, citecolor=blue,linkcolor=blue] {hyperref}
\usepackage{lineno}
\usepackage{bm}
\usepackage{rotating}
\usepackage[utf8]{inputenc}
\usepackage[separate-uncertainty=true,table-align-uncertainty=true,]{siunitx}

\let\oldequation\equation
\let\oldendequation\endequation
\renewenvironment{equation}
 {\linenomathNonumbers\oldequation}
 {\oldendequation\endlinenomath}

\begin{document}
\title{\boldmath Determination of Spin-parity Quantum Numbers for the Narrow Structure Near the $p\bar{\Lambda}$ Threshold in $e^+e^-\to pK^-\bar{\Lambda}+c.c.$}

\author{
M.~Ablikim$^{1}$, M.~N.~Achasov$^{13,b}$, P.~Adlarson$^{73}$, R.~Aliberti$^{34}$, A.~Amoroso$^{72A,72C}$, M.~R.~An$^{38}$, Q.~An$^{69,56}$, Y.~Bai$^{55}$, O.~Bakina$^{35}$, I.~Balossino$^{29A}$
% ...
\\
\vspace{0.2cm}
(BESIII Collaboration)\\
\vspace{0.2cm} {\it
	$^{1}$ Institute of High Energy Physics, Beijing 100049, People's Republic of China\\
% ...
$^{13}$ G.I. Budker Institute of Nuclear Physics SB RAS (BINP), Novosibirsk 630090, Russia\\
% ...
$^{29}$ INFN Sezione di Ferrara, (A)INFN Sezione di Ferrara, I-44122, Ferrara, Italy; (B)University of Ferrara,  I-44122, Ferrara, Italy\\
% ...
$^{34}$ Johannes Gutenberg University of Mainz, Johann-Joachim-Becher-Weg 45, D-55099 Mainz, Germany\\
	$^{35}$ Joint Institute for Nuclear Research, 141980 Dubna, Moscow region, Russia\\
% ...
$^{38}$ Liaoning Normal University, Dalian 116029, People's Republic of China\\
% ...
$^{55}$ Southeast University, Nanjing 211100, People's Republic of China\\
	$^{56}$ State Key Laboratory of Particle Detection and Electronics, Beijing 100049, Hefei 230026, People's Republic of China\\
% ...
$^{69}$ University of Science and Technology of China, Hefei 230026, People's Republic of China\\
% ...
$^{72}$ University of Turin and INFN, (A)University of Turin, I-10125, Turin, Italy; (B)University of Eastern Piedmont, I-15121, Alessandria, Italy; (C)INFN, I-10125, Turin, Italy\\
	$^{73}$ Uppsala University, Box 516, SE-75120 Uppsala, Sweden\\
% ...
$^{b}$ Also at the Novosibirsk State University, Novosibirsk, 630090, Russia\\
% ...
}
\vspace{0.4cm}
}

\begin{abstract}
A narrow structure in the $p\bar{\Lambda}$ system near the mass threshold,
named as $X(2085)$, is observed in the process $e^+e^- \to p K^-
\bar{\Lambda}$ with a statistical significance greater than $20\sigma$.
Its spin and parity are determined for the first time to be $J^P=1^+$
in an amplitude analysis, with statistical significance greater than
$5\sigma$ over other quantum numbers. The pole positions of $X(2085)$
are measured to be $M_{\rm pole}=(2086\pm4\pm6)$~MeV and $\Gamma_{\rm
  pole}=(56\pm5\pm16)$ MeV, where the first uncertainties are
statistical and the second ones are systematic. The analysis is based
on the study of the process $e^+e^-\to pK^-\bar{\Lambda}$ and uses the
data samples collected with the BESIII detector at the center-of-mass
energies $\sqrt{s}=4.008$, $4.178$, $4.226$, $4.258$, $4.416$, and
$4.682$ GeV with a total integrated luminosity of $8.35~\text{fb}^{-1}$.

\end{abstract}


\maketitle

Quantum chromodynamics~(QCD), the theory of the strong interaction, allows the existence of bound states beyond conventional mesons and baryons. Searching for these exotic states is one of the main interests in experimental hadron physics. An anomalous enhancement near the mass threshold of the $p\bar{\Lambda}$ system was first observed by the BES collaboration in the decay of $J/\psi\to pK^-\bar{\Lambda}$~\cite{BES_exp:Jpsi2pkLambda}. This enhancement is consistent with an $S$-wave Breit-Wigner function with a mass of $m=(2075\pm12(\text{stat.})\pm5(\text{syst.}))$ MeV and a constant width of $\Gamma=(90\pm35(\text{stat.})\pm9(\text{syst.}))$ MeV, but can also be described with a $P$-wave Breit-Wigner resonance. Therefore, the spin and parity of this structure were not determined. Similar evidence of a structure in $p\bar{\Lambda}$ was reported in several decays of $B$ mesons and charmonium states, such as $B^0\to p\bar{\Lambda}\pi^-$~\cite{BELLE_exp:pLambdapi}, $B^0\to \bar{p}\Lambda\pi^-$~\cite{BARAR_exp:pLambdapi}, $B^+\to p\bar{\Lambda}\gamma$~\cite{BELLE_exp:pLambdaGam}, $B^+\to p\bar{\Lambda}\pi^0$ ~\cite{BELLE_exp:pLambdaGampi},
$\psi(3686)\to pK^-\bar{\Lambda}$~\cite{BES_exp:Jpsi2pkLambda}, $\chi_{cJ}\to pK^-\bar{\Lambda}$~\cite{BES_exp:Chicj2pkLambda}, and their charge conjugations. However, the mass and width of the structure in $B$ decays were not determined. By replacing the $s$-quark with a $c$-quark, a similar structure is also observed in the $\bar{p}\Lambda^+_{c}$ system in the decay of $B^-\to \Lambda^+_{c}\bar{p}\pi^-$~\cite{BELLE_exp:B2LambdaCppi}.

Theoretically, the near-threshold enhancement in the $p\bar{\Lambda}$ system was investigated in scenarios of $q^3\bar{q}^3$ meson~\cite{Theory:q3qbar3_meson}, baryon-antibaryon SU(3) nonets~\cite{Theory:Baryon_antiBaryon_nonets}, final state interaction~\cite{Theory:pLambda_fsi}, or chiral effective field theory~\cite{Theory:Chiral_effective_pLambda1}. The quantum numbers $J^P=0^-$ are excluded by combining $C$ and $P$ symmetries with SU(3) flavor symmetry~\cite{Theory:Quantum_number_0_pm}. The hypothesis of an $S$-wave $p\bar\Lambda$ bound state was rejected in the quark model by considering annihilation interaction~\cite{Theory:pLambda_s_wave}. The mass spectrum of $p\bar\Lambda$ in $B$ meson decays was not fully understood in the naive factorization picture with limited statistics \cite{Chua,Geng}. Further experimental studies of the near threshold enhancement in the $p\bar{\Lambda}$ system in various processes are critical for validating different theoretical models.

In this Letter, we report the observation of a narrow structure near the $p\bar{\Lambda}$ mass threshold, named as $X(2085)$, in the process $e^+e^-\to pK^-\bar{\Lambda}$ at the center-of-mass energies $\sqrt{s}=4.008$, $4.178$, $4.226$, $4.258$, $4.416$, and $4.682$ GeV. The quantum numbers and resonance parameters of $X(2085)$ are determined from an amplitude analysis. A detailed measurement of the energy dependence of the $e^+e^-\to pK^-\bar{\Lambda}$ cross sections can be found in a separate paper~\cite{BESIII_exp:ee2pkl}. Throughout this Letter, charged conjugated modes are always implied.

The BESIII detector is a magnetic spectrometer~\cite{Ablikim:2009aa}
located at the Beijing Electron Positron Collider~(BEPCII)~\cite{Yu:IPAC2016-TUYA01}.
The cylindrical core of the BESIII detector consists of a main drift chamber filled with a helium-based gas~(MDC), a plastic scintillator time-of-flight
system~(TOF), and a CsI(Tl) electromagnetic calorimeter~(EMC), which are all
enclosed in a superconducting solenoidal magnet providing a 1.0~T magnetic
field. The flux-return yoke is instrumented with resistive plate
chambers arranged in 9 layers in the barrel and 8 layers in the endcaps for muon identification. The
acceptance for charged particles and photons is 93\% of $4\pi$ solid angle.
The charged-particle momentum resolution at 1.0~GeV/$c$ is $0.5\%$, and the
specific energy loss resolution is $6\%$ for electrons from
Bhabha scattering. The EMC measures photon energies with a resolution of
$2.5\%$~($5\%$) at $1$~GeV in the barrel (end cap) region. The time resolution
of the TOF barrel part is 68~ps, while that of the end cap part is 110~ps. The end cap TOF system was upgraded in 2015 with multi-gap resistive plate chamber technology, providing a time resolution of 60 ps~\cite{TOF_upgrade}.


Monte-Carlo (MC) simulated samples produced with a {\sc geant4}-based~\cite{GEANT4}
software package, which includes the geometric description of
the BESIII detector and the detector response, are used to determine
the detection efficiencies and to estimate the backgrounds. The simulation includes the beam energy spread and initial state radiation in the $e^+e^-$ annihilations modeled with the {\sc kkmc} generator~\cite{KKMC}. The inclusive MC sample consists of the production of open-charm processes, the initial state radiation of vector charmonium(-like) states, and the continuum process incorporated in {\sc kkmc}~\cite{KKMC}. The known decay states are modeled with {\sc besevtgen}~\cite{evtgen} using branching fractions taken from the Particle Data Group~\cite{pdg2022}, and the remaining unknown decays from the charmonium states with {\sc lundcharm}~\cite{LUNDCHARM1}. Final state radiation from charged final state particles is incorporated with {\sc photos}~\cite{PHOTOS} package.

The candidates for $e^+e^-\to pK^-\bar{\Lambda}$ are required to have four charged tracks with zero net charge. Charged tracks are required to start from the region $|d_z|<$ 20 cm and $|\cos\theta|<$ 0.93, where $|d_z|$ is the distance of closest approach to the interaction point~(IP) along the $z$-axis, and $\theta$ is the polar angle relative to the $z$-axis, which is taken to be the symmetry axis of the MDC.

To reconstruct $\Lambda(\bar{\Lambda})$, all possible track pairs with opposite charges are assigned as $p\pi^-(\bar{p}\pi^+)$. The $p\pi^-(\bar{p}\pi^+)$ trajectories are constrained to originate from a common vertex by applying a vertex fit, the $\chi^2$ of which is required to be less than 100. The decay length of any $\Lambda(\bar{\Lambda})$ candidate must be greater than twice the standard deviation of the vertex resolution.  The invariant mass of $p\pi^-(\bar{p}\pi^+)$ is required to be within the $\Lambda(\bar{\Lambda})$ signal mass window of $|M_{p\pi^-(\bar{p}\pi^+)}-M_{\Lambda}|< 6$\,MeV/$c^2$. In each event, only exactly one $\Lambda(\bar{\Lambda})$ candidate is allowed. The other two charged tracks are assigned according to their charges as proton and kaon not from $\Lambda(\bar{\Lambda})$ decays. To ensure that these tracks originate from the IP, an additional requirements of $|d_z|< 10$~cm and $|d_r|< 1$~cm is imposed, where $|d_r|$ is the distance of closest approach to the interaction point in the transverse plane.
A four-constraint~(4C) kinematic fit is imposed on the selected charged particles under the hypothesis $e^+e^-\to pK^-\bar{\Lambda}$. The $\chi^2$ of the 4C kinematic fit is required to be less than 100.

The background from $e^+e^-\to(\gamma)e^+e^-$ is suppressed by rejecting events with $|\cos\theta_{K}|>0.83$, where $\theta_{K}$ is the polar angle of the kaon candidate.

\begin{figure}[htbp]
	\centering
	\includegraphics[width=\columnwidth]{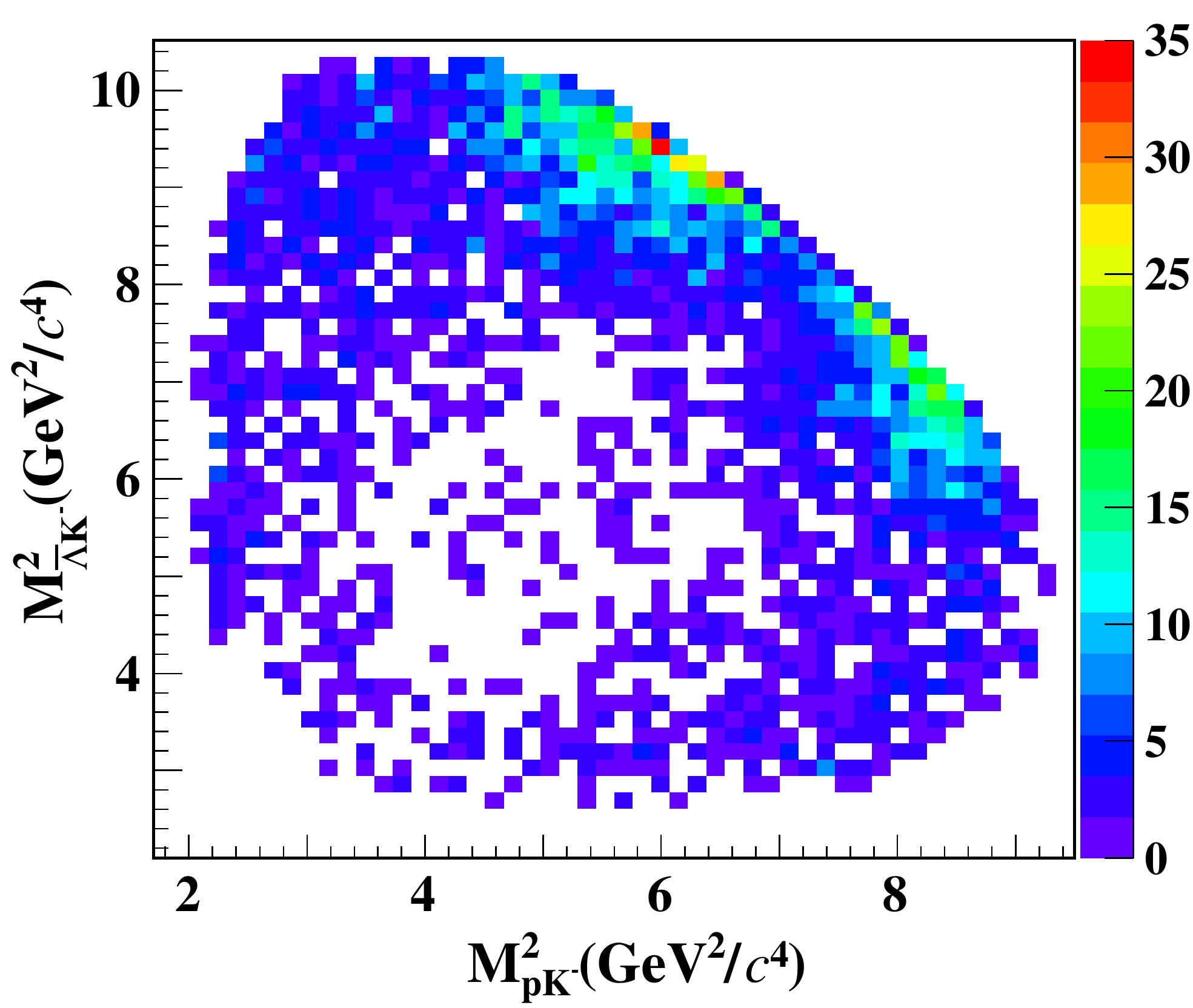}
	\caption{Dalitz plot of the selected $e^+e^-\to pK^-\bar{\Lambda}$ events in data at $\sqrt{s}=4.178$ GeV.}
	\label{fig:Dalitz}
\end{figure}

After applying all the above selection criteria, a total of 3883 candidate events for $e^+e^-\to pK^-\bar{\Lambda}$ survives in the data sample at $\sqrt{s}=4.178$ GeV, which is the largest data set under consideration. A study of the inclusive MC leads to an estimated background yield of 52 events. Figure~\ref{fig:Dalitz} shows the Dalitz plot of the selected events, where a significant near-threshold enhancement in the $p\bar{\Lambda}$ system is observed. In addition, contributions of some excited $\Lambda$ and nucleon states can also be seen.

The structures are investigated via an amplitude analysis, in which the amplitudes for the sequential processes $e^+e^-\to\gamma^*\to X^+ K^- (X^+\to p\bar{\Lambda})$, $e^+e^-\to\gamma^*\to N^{*+} \bar{p}(N^{*+}\to K^+\bar{\Lambda})$, $e^+e^-\to\gamma^*\to \Lambda^{*} \bar{\Lambda}(\Lambda^{*}\to pK^-)$, and their charge conjugates, are constructed using the relativistic covariant tensor amplitude formalism~\cite{FDC}. 
These effective vertices $\Gamma$ are deduced from an effective Lagrangian by considering $C$- and $P$-parity invariance, Lorentz invariance, and $CPT$ invariance.

The amplitude of a process containing a specific resonance is written as
\begin{equation}
\mathcal{A}_j=\epsilon^{*\alpha}(p_0,m)\bar u(p_1,s_1)\Gamma_{1\alpha\mu_1\mu_2...}\Gamma_2^{\mu_1\mu_2...}v(p_2,s_2)BW(s),
\end{equation}
where $\epsilon^{*}$ is the $\gamma^*$ polarization vector; $u(p_1,s_1)$ and $v(p_2,s_2)$ are the free Dirac spinors for proton and $\bar\Lambda$, respectively; $\Gamma_1$ and $\Gamma_2$ are the strong interaction vertices describing the resonance couplings with $\gamma^*$-$p$ and $\bar{\Lambda}$-$K^-$, respectively; $BW(s)$ is a Breit-Wigner function.

The lineshape of $X(2085)$ is parametrized by a relativistic Breit-Wigner function
\begin{equation}
BW(s)=\frac{1}{m_0^2-m^2-im_0\Gamma(m)},
\end{equation}
with a mass dependent width given by
\begin{equation}
\Gamma(m)=\Gamma_{0}\left(\frac{q}{q_0}\right)^{2l+1}\frac{m_0}{m}B^2_l(q,q_0,d).
\end{equation}
Here, $m$ is the invariant mass of the $p\bar{\Lambda}$ system; $m_0$ and $\Gamma_0$ are the mass and width of $X(2085)$, respectively; $q(q_0)$ is the three-momentum of the proton in the rest frame of $p\bar{\Lambda}$, which is calculated with the invariant mass $m(m_0)$; $l$ is the orbital angular momentum of the $p\bar{\Lambda}$ system; $B_l(q,q_0,d)$ is the reduced Blatt-Weisskopf barrier factor~\cite{Barrier_factor}; $d$ is the radius of the centrifugal barrier chosen as $d=0.73$~fm~\cite{BES_exp:psip2kketa}, corresponding to the range of the strong interaction in hadronic decays. The other intermediate states are described with a constant width, \emph{i.e.} $\Gamma(m)=\Gamma_{0}$.

\begin{figure*}[htbp]
	\centering
	\includegraphics[width=\textwidth]{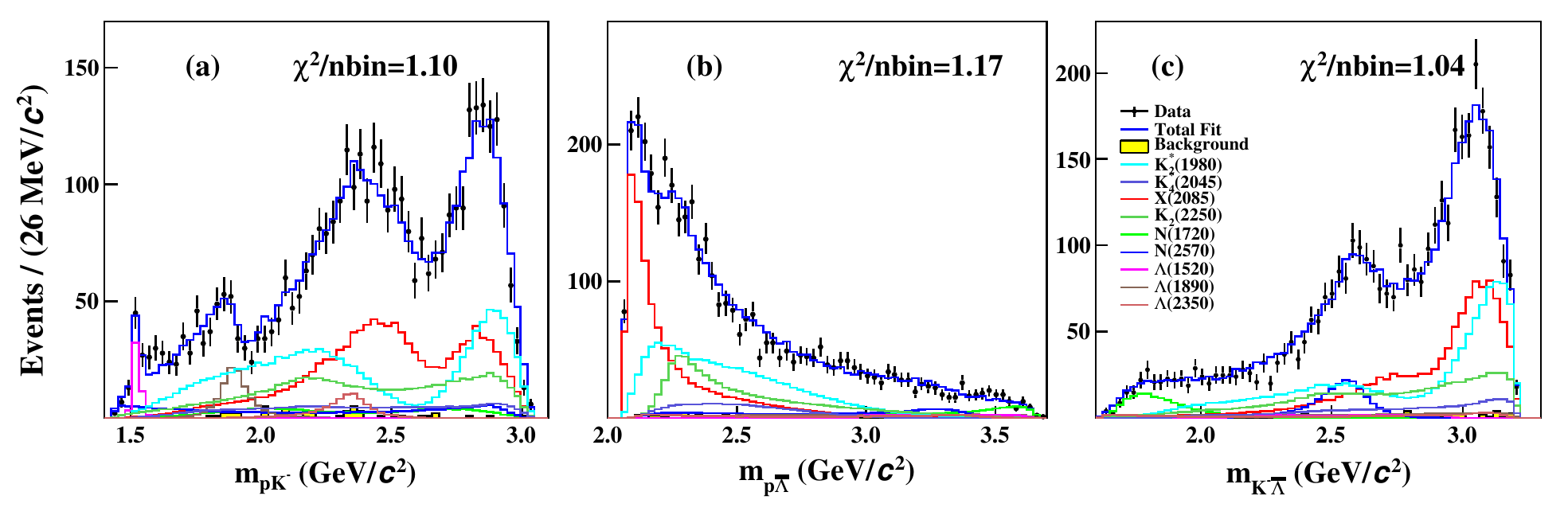}
	\caption{Fit projections of the amplitude fit result on the (a) $pK^-$, (b) $p\bar{\Lambda}$, and (c) $K^-\bar{\Lambda}$ invariant mass distributions of the $e^+e^-\to pK^-\Lambda$ candidate events in data taken at $\sqrt{s}=4.178$ GeV. The points with errors are data, the blue curve is the fit result, the curves in various colors denote different resonant components and the yellow filled histogram is the simulated background.}
	\label{fig:PWA_fit}
\end{figure*}

The complex coupling constants of the amplitudes are determined by an unbinned maximum likelihood fit using {\sc minuit}~\cite{Minuit}. The background contribution is estimated with the inclusive MC sample and subtracted from the likelihood.
The amplitude analysis is first performed to the data taken at $\sqrt{s}=4.178$ GeV, which has the largest luminosity among the six data sets. The contributions from $X(2085)$ as well as from other excited kaon, $\Lambda$, and nucleon states are evaluated. The baseline solution is obtained with only amplitudes having a statistical significance greater than 5 $\sigma$, including $X(2085)$, $K^*_2(1980),K^*_4(2045),K_2(2250),\Lambda(1520),\Lambda(1890),\Lambda(2350),$ $N(1720)$, and $N(2570)$. The resonance parameters of excited kaon, $\Lambda$, and nucleon states are fixed to individual world average values~\cite{pdg2022}. Possible $J^P$ assignments, $0^-$, $1^-$, $1^+$, $2^+$, and $2^-$, for $X(2085)$ are tested with its mass and width as free parameters, as shown in Table~\ref{tab:Compare_JP}.
The statistical significance of $J^P=1^+$ over the other four quantum numbers is estimated by following the method in Ref.~\cite{Exp:Lambdab2JpsiKp}. When discriminating between different $J^P$ assignments to the $X(2085)$, the assumption of a $\chi^2$ distribution allows the calculation of a lower limit on the significance of its rejection~\cite{Theory:Quantum_significance}, using the change in the $\Delta(-2\ln\mathcal{L})=2(\ln\mathcal{L}_{1^+}-\ln\mathcal{L}_{J^P})$ versus the change of 1 degree of freedom under the disfavored $J^P$ hypothesis. The log-likelihood difference is determined with the fit to data. Compared to the $J^P=1^+$ hypothesis, the $J^P=0^-,1^-,2^+,$ and $2^-$ hypotheses are rejected with statistical significances of 9.3$\sigma$, 8.1$\sigma$, 9.8$\sigma$, and 5.7$\sigma$, respectively, as summarized in Table ~\ref{tab:Compare_JP}.  The $X(2085)$ is observed with statistical significance greater than 20$\sigma$. Since the mass and width of a mass-dependent-width BW are model dependent and may differ from the actual resonance properties, we solve for $P = M_{\rm pole}-i\Gamma_{\rm pole}/2$, the position in the complex ($M$, $\Gamma$) plane where the BW denominator is zero, and use $M_{\rm pole}$ and $\Gamma_{\rm pole}$ to characterize the mass and width of the $X(2085)$ resonance. The pole parameters of $X(2085)$ are determined to be $M_{\rm pole}=(2085\pm6)$~MeV and $\Gamma_{\rm pole}=(62\pm10)$ MeV. 
The amplitude fit results are consistent with data in the $M_{pK^-}$, $M_{p\bar{\Lambda}}$, and $M_{K^-\bar{\Lambda}}$ distributions, as shown in Fig.~\ref{fig:PWA_fit}.

 \begin{table}[htbp]
	\centering
	\setlength\tabcolsep{6pt}
	\caption{Resultant resonance parameters of $X(2085)$ and estimated significances under various hypotheses. The uncertainties are only statistical.}
	\begin{tabular}{l|SSc}
		\hline
		\hline
		$J^P$ & {$M_{\rm pole}$~(MeV)} & {$\Gamma_{\rm pole}$~(MeV)} & $\sqrt{\Delta(-2\ln\mathcal{L})}$ \\
		\hline
		$1^+$  &  2085 +- 6  & 62 +- 10 & ---  \\
		$0^-$  &  2095 +- 9  & 140 +- 6& 9.3  \\
		$1^-$  &  2063 +- 12  & 130 +- 9 & 8.1  \\
		$2^+$  &  2077 +- 2  & 122 +- 8 & 9.8  \\
		$2^-$  &  2082 +- 16  & 46 +- 36 & 5.7 \\
		\hline
		\hline
	\end{tabular}
	\label{tab:Compare_JP}
\end{table}

The fit with the same set of amplitudes is performed to data sets taken at the other energy points. The obtained results are summarized in Table~\ref{tab:other_points}.  The pole positions of $X(2085)$ are stable and independent of $\sqrt{s}$. The averaged pole parameters are determined to be $M_{\rm pole}=(2086\pm4)$ MeV and $\Gamma_{\rm pole}=(56\pm5)$ MeV.

 \begin{table}[htbp]
	\centering
	\setlength\tabcolsep{7pt}
	\caption{The integrated luminosity $\mathcal{L}_{\rm int}$~(in
          pb$^{-1}$), year of data taking, and obtained pole
          parameters of $X(2085)$~(in MeV) at various energy points
          $\sqrt{s}$~(in GeV). The uncertainties are statistical
          only.}
        \sisetup{table-figures-uncertainty=3}
	\begin{tabular}{l|S[table-format=4.1,table-figures-uncertainty=2]c
          S[table-format=4.0,table-figures-uncertainty=2]
          S[table-format=2.0,table-figures-uncertainty=1]}
		\hline
		\hline 
		$\sqrt{s}$  & \multicolumn{1}{c}{$\mathcal{L}_{\rm int}$} &  Year & {$M_{\rm pole}$} & $\Gamma_{\rm pole}$ \\
		\hline
		$4.008$  & 482.0 +- 4.7 & 2011 & 2085 +- 14  & 50 +- 16 \\
		$4.178$  & 3189.0 +- 31.9 & 2016 & 2085 +- 6  &  62 +- 10 \\
		$4.226$  & 1100.9 +- 7.0 & 2013 & 2088 +- 10  & 68 +- 12 \\
		$4.258$  & 828.4 +- 5.5 & 2013 & 2083 +- 11  & 48 +- 10 \\
		$4.416$  & 1090.7 +- 7.2 & 2014 & 2088 +- 13  & 56 +- 12 \\
		$4.682$  & 1669.3 +- 9.0 & 2020 & 2092 +- 10  & 54 +- 10 \\
		\hline
		Average  & {---}& {---}& 2086 +- 4  & 56 +- 5 \\
		\hline
		\hline
	\end{tabular}
	\label{tab:other_points}
\end{table}

The systematic uncertainties on the resonance parameters of $X(2085)$ include the quoted resonance parameters, the $\cos\theta_{K}$ requirement, the $\Lambda(\bar{\Lambda})$ signal mass window, background estimation, and the mass resolution, as summarized in Table~\ref{tab:sys_X2086}. The uncertainties, except those associated with the background estimation, are considered to be correlated among various energy points and studied at $\sqrt{s}=4.178$ GeV. The systematic uncertainty arising from taking the radius $d=0.73$~fm is estimated with alternative radii 0.39~fm and 1.32~fm~\cite{BES_exp:psip2kketa}. The systematic uncertainty related to possible contributions from excited $\Sigma$ states is estimated by replacing the $\Lambda^*$ states in the nominal solution with similar $\Sigma^*$ states. The systematic uncertainty due to quoted resonance parameters of the other seven resonances is estimated by sampling their parameters according to their uncertainties, repeating the fit, and taking the width of the resulting distribution as systematic uncertainty. The systematic uncertainty of the requirement $|\cos\theta_{K}|< 0.83$ is estimated by enlarging or shrinking this requirement by 0.01. The maximum change between resonance parameters of $X(2085)$ is taken to be the corresponding uncertainty. The systematic uncertainty related to the $\Lambda(\bar{\Lambda})$ signal mass window is estimated by varying it by 1 MeV.  The systematic uncertainty due to the background estimation is evaluated by performing another amplitude fit without considering the background contribution. For each of these sources, the largest change of the $X(2085)$ resonance parameters is assigned as the systematic uncertainty. The systematic uncertainty due to the mass resolution is estimated with a simulation study as described in Ref.~\cite{mass_resolution_sys}. This study is used to derive a relation between the measured pole parameters and their true values, and the difference between the parameters from the fit to data and the derived true values is taken as the corresponding systematic uncertainty.

 \begin{table}[htbp]
	\centering
	\caption{Systematic uncertainties on the pole positions of $X(2085)$.}
	\begin{tabular}{lcc}
		\hline
		\hline
		Source &  $M_{\rm pole}$~(MeV)& $\Gamma_{\rm pole}$~(MeV)\\
		\hline
		Radius $d$  & 4.8 & 15.2 \\
		Excited $\Sigma$ states & 2.7 & 4.8 \\		
		Resonance parameters  & 0.8 & 1.7 \\
		$|\cos\theta_{K}|$ requirement & 0.4 & 0.2\\
		$\Lambda(\bar{\Lambda})$ signal mass window & 0.8 & 1.2 \\
		Background estimation & 1.3 & 2.0 \\
		Mass resolution &  0.3 & 0.2\\
		\hline
		Total& 5.8 &  16.2 \\
		\hline
		\hline
	\end{tabular}
	\label{tab:sys_X2086}
\end{table}

In summary, with $8.35$ fb$^{-1}$ of $e^+e^-$ collision data taken at $\sqrt{s}=4.008$, $4.178$, $4.226$, $4.258$, $4.416$ and $4.682$ GeV, the process $e^+e^-\to pK^-\bar{\Lambda}$ is studied. The enhancement near the mass threshold of $p\bar{\Lambda}$ system, $X(2085)$, is observed with a statistical significance greater than $20\sigma$. 
 The spin and parity of $X(2085)$ are determined with
an amplitude analysis to be $1^+$ with a statistical significance greater than 5$\sigma$ over other quantum numbers. Under the assignment of $J^P=1^+$, the pole positions of $X(2085)$ are measured to be $M_{\rm pole}=(2086\pm4\pm6)$ MeV and $\Gamma_{\rm pole}=(56\pm5\pm16)$ MeV, respectively. They match neither any known excited kaon state observed in experiments~\cite{pdg2022} nor any state predicted by the potential model~\cite{Theory:Kaon_spectrum}. The anomalous narrow width and the mass near the $p\bar{\Lambda}$ threshold may suggest exotic properties of $X(2085)$. No conclusion can be drawn that $X(2085)$ is the same structure as the one observed in Ref.~\cite{BES_exp:Jpsi2pkLambda} since limited information was given.
Further studies about the properties of $X(2085)$ in the decays of $J/\psi\to pK^-\bar{\Lambda}$ and $\psi(3686)\to pK^-\bar{\Lambda}$ with larger $J/\psi$ and $\psi(3686)$ data samples~\cite{white_book} are desirable to provide more precise information. Searching for its isospin partner in $e^+e^-\to nK_S\bar{\Lambda}$ final state is crucial to check if $X(2085)$ has an isospin of 1/2.

The BESIII collaboration thanks the staff of BEPCII and the IHEP computing center for their strong support. This work is supported in part by National Key R\&D Program of China under Contracts Nos. 2020YFA0406300, 2020YFA0406400; National Natural Science Foundation of China (NSFC) under Contracts Nos. 12175244, 12035009, 11875170, 11875262, 11565006, 11505034, 11635010, 11735014, 11835012, 11935015, 11935016, 11935018, 11961141012, 12022510, 12025502, 12035013, 12061131003, 12192260, 12192261, 12192262, 12192263, 12192264, 12192265; the Chinese Academy of Sciences (CAS) Large-Scale Scientific Facility Program; the CAS Center for Excellence in Particle Physics (CCEPP); Joint Large-Scale Scientific Facility Funds of the NSFC and CAS under Contract No. U1832207; CAS Key Research Program of Frontier Sciences under Contracts Nos. QYZDJ-SSW-SLH003, QYZDJ-SSW-SLH040; 100 Talents Program of CAS; The Institute of Nuclear and Particle Physics (INPAC) and Shanghai Key Laboratory for Particle Physics and Cosmology; ERC under Contract No. 758462; European Union's Horizon 2020 research and innovation programme under Marie Sklodowska-Curie grant agreement under Contract No. 894790; German Research Foundation DFG under Contracts Nos. 443159800, 455635585, Collaborative Research Center CRC 1044, FOR5327, GRK 2149; Istituto Nazionale di Fisica Nucleare, Italy; Ministry of Development of Turkey under Contract No. DPT2006K-120470; National Research Foundation of Korea under Contract No. NRF-2022R1A2C1092335; National Science and Technology fund; National Science Research and Innovation Fund (NSRF) via the Program Management Unit for Human Resources \& Institutional Development, Research and Innovation under Contract No. B16F640076; Polish National Science Centre under Contract No. 2019/35/O/ST2/02907; Suranaree University of Technology (SUT), Thailand Science Research and Innovation (TSRI), and National Science Research and Innovation Fund (NSRF) under Contract No. 160355; The Royal Society, UK under Contract No. DH160214; The Swedish Research Council; U. S. Department of Energy under Contract No. DE-FG02-05ER41374.

\end{document}